\documentclass[]{cupconf}
\input{psfig}

  \checkfont{eurm10}
  \iffontfound
    \IfFileExists{upmath.sty}
      {\typeout{^^JFound AMS Euler Roman fonts on the system,
                   using the 'upmath' package.^^J}%
       \usepackage{upmath}}
      {\typeout{^^JFound AMS Euler Roman fonts on the system, but you
                   dont seem to have the}%
       \typeout{'upmath' package installed. cupconf.cls can take advantage
                 of these fonts,^^Jif you use 'upmath' package.^^J}%
      }
  \else
  \fi

  \checkfont{msam10}
  \iffontfound
    \IfFileExists{amssymb.sty}
      {\typeout{^^JFound AMS Symbol fonts on the system, using the
                'amssymb' package.^^J}%
       \usepackage{amssymb}%

      }{}
  \fi

  \IfFileExists{amsbsy.sty}
    {\typeout{^^JFound the 'amsbsy' package on the system, using it.^^J}%
     \usepackage{amsbsy}}
    {}

\newsavebox{\astrutbox}
\sbox{\astrutbox}{\rule[-5pt]{0pt}{20pt}}

\newcommand\etal{\mbox{\textit{et al.}}}

\def\Ha{H$\alpha$}
\def\Lha{\cal L}
\def\hii{{\sc H\thinspace ii}}
\def\hi{{\sc H\thinspace i}}
\def\ergs{{\rm\,erg\,s^{-1}}}
\def\cc{{\rm\,cm^{-3}}}

\title[Massive Star Feedback]
{The Local Group as an Astrophysical Laboratory for Massive Star Feedback}

\author[M. S. Oey]%
{M.\ns S.\ns O\ls E\ls Y$^1$}
\affiliation{$^1${Lowell Observatory, 1400 W. Mars Hill Rd., Flagstaff, 
	AZ\ \ \ 86001, USA}}

\pubyear{2003}

\begin{document}

\maketitle

\begin{abstract}
The feedback effects of massive stars on their galactic and
intergalactic environments can dominate evolutionary processes in
galaxies and affect cosmic structure in the Universe.  Only the Local
Group offers the spatial resolution to quantitatively study feedback
processes on a variety of scales.  Lyman continuum
radiation from hot, luminous stars ionizes \hii\ regions and is
believed to dominate production of the warm component of the
interstellar medium (ISM).  Some of this radiation apparently escapes from
galaxies into the intergalactic environment.  Supernovae and strong
stellar winds generate shell structures such as supernova remnants,
stellar wind bubbles, and superbubbles around OB associations.  Hot
($10^6$ K) gas is generated within these shells, and is believed to be
the origin of the hot component of the ISM.  Superbubble activity thus
is likely to dominate the ISM structure, kinematics, and phase balance
in star-forming galaxies.  Galactic superwinds in starburst galaxies
enable the escape of mass, ionizing radiation, and heavy elements.
Although many important issues remain to be resolved, there is little
doubt that feedback processes plays a fundamental role in energy cycles on
scales ranging from individual stars to cosmic structure.  This
contribution reviews studies of radiative and mechanical feedback in
the Local Group. 
\end{abstract}

\firstsection 
\section{Introduction}

The Local Group is especially suited as a laboratory for studying the
effects of the massive star population on the galactic environment.
There are three types of massive star feedback:
1. Radiative feedback, i.e., ionizing emission, which results in
photoionized nebulae and diffuse, warm ($10^4$ K) ionized gas;
2. Mechanical feedback, predominantly from supernovae (SNe), resulting
in supernova remnants (SNRs), superbubbles, and galactic superwinds;
and 3. Chemical feedback from nucleosynthesis in SNe and massive star
evolution, which drives galactic chemical evolution.  Since the last
will be reviewed by Don Garnett and Monica Tosi in this volume, I
will address here only the radiative and mechanical feedback processes.

Only in the Local Group can we spatially resolve the
the various physical parameters that determine, and result from, the
interaction of massive stars with their immediate environment.
Radiative and mechanical feedback return large quantities of
energy to the galactic environment, both with luminosities $\log L$ of
order $36 - 41\ \ergs$ for typical OB associations.  These processes
may dominate the balance between different temperature phases of the
interstellar medium (ISM), profoundly affecting the structure and
kinematics of the ISM, star formation, and other evolutionary
processes in star-forming galaxies.  

\section{Radiative Feedback}

Photoionization results in \hii\ regions and ionized gas, which are
especially familiar and photogenic in the Milky Way and Magellanic
Clouds. 

\subsection{Nebular emission-line diagnostics}

The nebular emission-line spectra offer vital
diagnostics of conditions in these regions, and are especially
powerful if modeled with tailored photoionization models.  However, a
class of ``semi-empirical'' line diagnostics are widely used to probe
nebular parameters.  For example, the parameter (V\'\i lchez \& Pagel
1988) 
\begin{equation}\label{etap}
\eta^\prime \equiv \rm \frac{[O\thinspace II]\lambda3727/
	[O\thinspace III]\lambda\lambda4959,5007}
	{[S\thinspace II]\lambda6724/
	[S\thinspace III]\lambda\lambda9069,9532}
\end{equation}
and [Ne III]/H$\beta$ (Oey et al. 2000) probe the ionizing stellar
effective temperature, while the parameters (Pagel et al. 1979)
\begin{equation}\label{eqR23}
R23 \equiv \rm \bigl([O\thinspace II]\lambda3727 + 
	[O\thinspace III]\lambda\lambda4959,5007 \bigr) / {\rm H\beta}
\end{equation}  
and (V\'\i lchez \& Esteban 1996; Christensen et al. 1997)
\begin{equation}\label{eqS23}
S23 \equiv \rm \bigl([S\thinspace II]\lambda6724 + 
	[S\thinspace III]\lambda\lambda9069,9532 \bigr) / {\rm H\beta}
\end{equation}  
estimate the oxygen and sulfur abundances.  Using observations of LMC
\hii\ regions, Oey \& Shields (2000) question the reliability of $S23$
and show that 
\begin{equation}\label{eqS234}
S234 \equiv \rm \bigl([S\thinspace II]\lambda6724 + 
	[S\thinspace III]\lambda\lambda9069,9532 +
	[S\thinspace IV]10.5\mu \bigr) / {\rm H\beta}
	\quad ,
\end{equation}  
a superior diagnostic of sulfur abundance, can be easily estimated
from optical line strengths.  All of these semi-empirical diagnostics
must first be similarly tested and calibrated using objects with 
independently constrained parameters.  The Local Group offers by far
the best nebular samples in which to simultaneously:  obtain spectral
classifications of the ionizing stars; evaluate the gas morphology
relative to the ionizing stars, thereby constraining the nebular
ionization parameter; and estimate the abundances.  These
are the three primary parameters that determine the nebular emission.
Thus, having empirical constraints on photoionization models, the
behavior of the emission-line diagnostics can be calibrated.
The nebular emission can also be used to constrain the hot stellar
atmosphere models themselves, since these NLTE, expanding atmospheres
are complicated and difficult to model.  Such studies have been
carried out for O and WR stars in the Magellanic Clouds and the Galaxy
(Oey et al. 2000; Kennicutt et al. 2000; Crowther et al. 1999).  

Another important test is to evaluate the degree to which \hii\
regions are indeed radiation-bounded, as is normally assumed; the
escape of ionizing radiation is a critical question for the ionization
of the diffuse, warm ionized medium (see below), as well as the use of H
recombination emission as a star formation tracer.  The escape of
ionizing radiation from the host galaxies themselves is also vital to
understanding the ionization state of the intergalactic medium (IGM)
and the reionization of the early Universe.  We can test whether \hii\
regions are radiation-bounded by simply comparing the observed stellar
spectral types, and thus inferred ionizing flux, with the observed nebular
emission.  Currently, it appears that while most nebulae are
radiation-bounded, a large subset apparently are density-bounded (Oey
\& Kennicutt 1997; Hunter \& Massey 1990).

With adequate calibrations of nebular diagnostics against Local Group
objects, we can infer various physical parameters for distant,
unresolved star-forming regions. 

\subsection{Statistical properties of \hii\ regions}

\begin{figure*}
\psfig{figure=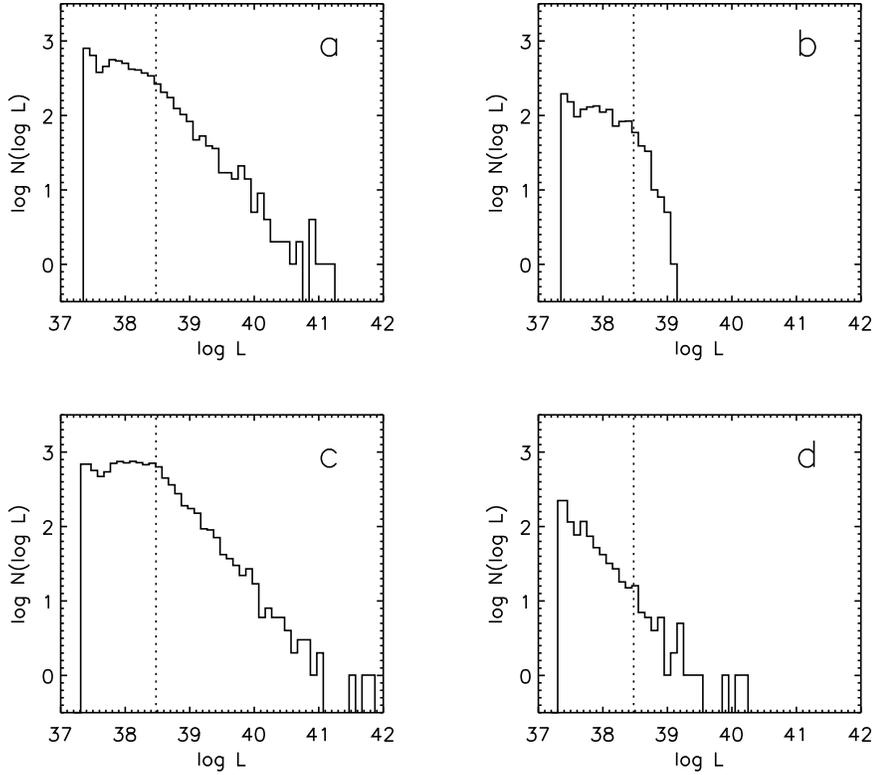,width=5.0truein}
\vspace*{-2.0truein}
\caption{Monte Carlo models of the \hii\ LF from Oey \& Clarke (1998a):
($a$)  constant nebular creation rate for a full power law in $N_*$
given by equation~\ref{eq_N*}; ($b$) continuous creation with an
upper cutoff in $N_*$ of 10 ionizing stars; ($c$) zero-age
instantaneous burst for all objects, with a full power law in $N_*$;
($d$) the evolved burst in ($c$) after 7 Myr.  The maximum stellar
ionizing luminosity in the IMF here corresponds to log ${\Lha} = 38.5$
(vertical dotted line).
\label{F_hlf}}
\end{figure*}

Statistical properties of \hii\ region populations offer important
quantitative characterizations of global star formation in galaxies.
The \hii\ region luminosity function (\hii\ LF) reveals the relative
importance of major star-forming events, hosting super star clusters,
and smaller, ordinary OB associations.  The \hii\ LF has been
determined for many nearby galaxies, including all of the star-forming
galaxies in the Local Group (Milky Way: Smith \& Kennicutt 1989, McKee
\& Williams 1997; Magellanic Clouds, M31, M33:  Kennicutt et al. 1989,
Walterbos \& Braun 1992, Hodge et al. 1999; IC 10, Leo A, Sex A, Sex B, GR8,
Peg, WLM:  Youngblood \& Hunter 1999).  There is agreement that the
\hii\ LF universally appears to be described by a power law:
\begin{equation}\label{eq_hiilf}
N({\Lha})\ d{\Lha} \propto {\Lha}^{-a}\ d{\Lha} \quad ,
\end{equation}
with a power-law index $a\sim 2$ for the differential LF.
Figure~\ref{F_hlf} presents Monte Carlo
models by Oey \& Clarke (1998a) that show the existence of
a flatter slope below log ${\Lha} \sim 37.5 - 38.5$, owing to a transition
at low luminosity to objects dominated by small number statistics in
the ionizing stellar population.  We also see that
the \hii\ LF can offer some insights on the nature and history of the
very most recent global star formation, within the last $\sim10$ Myr.

The observed behavior of the \hii\ LF is consistent with the existence
of a universal power law for the number of ionizing stars $N_*$ per
cluster (Oey \& Clarke 1998a):
\begin{equation}\label{eq_N*}
N(N_*)\ dN_* \propto N_*^{-2}\ dN_* \quad .
\end{equation}
This is consistent with direct observations of the cluster luminosity
and mass functions in a variety of regimes (see Chandar, this volume;
Elmegreen \& Efremov 1997; Meurer et al. 1995; Harris \& Pudritz
1994).  Such a universal power-law for the cluster mass function is
fundamental, similar to the stellar initial mass function (IMF; e.g.,
Oey \& Mu\~noz-Tu\~non 2003). 

The nebular size distribution has also been determined in many
galaxies, although it has been studied in less detail than the \hii\
LF (Milky Way, Magellanic Clouds, M31, M33, NGC 6822:  van den Bergh
1981, Hodge et al. 1999; IC~10, Leo A, Sex A, Sex B, GR8, Peg, WLM:
Hodge 1983, Youngblood \& Hunter 1999).  In his pioneering work, van
den Bergh (1981) described the size distribution as an exponential:
\begin{equation}\label{eq_sizeexp}
N_R \propto {\rm e}^{-R/R_0} \quad .
\end{equation}
However, this relation is difficult to reconcile with the power-law
form of the \hii\ LF.  To first order, the \Ha\ luminosity ${\Lha}$
should scale with the 
volume emission as $R^3$, and thus the size distribution should have a
similar power-law form to that of the \hii\ LF, but with exponent
$b = 2-3a$.  We can see in Figure~\ref{F_hlf}$a$ and $c$ that the
existence of the turnover in the \hii\ LF described above can cause
the entire \hii\ LF to mimic an exponential form.  Thus we suggest
that the intrinsic form of the size distribution is also a power-law
relation: 
\begin{equation}\label{eq_sizepow}
N(R)\ dR \propto R^{-b}\ dR \quad .
\end{equation}
With a slope also flattening below a value of log $R\sim 130$ pc,
corresponding to the transition in the \hii\ LF above, this form of the
size distribution is also a good description of the available data
(Oey et al. 2003).  Our initial investigation shows good agreement
between observations and the predicted value for $b = 4$,
implied by the \hii\ LF slope $a = 2$.

\subsection{Warm Ionized Medium}

\begin{figure*}
\hspace*{0.1 truein}
\psfig{figure=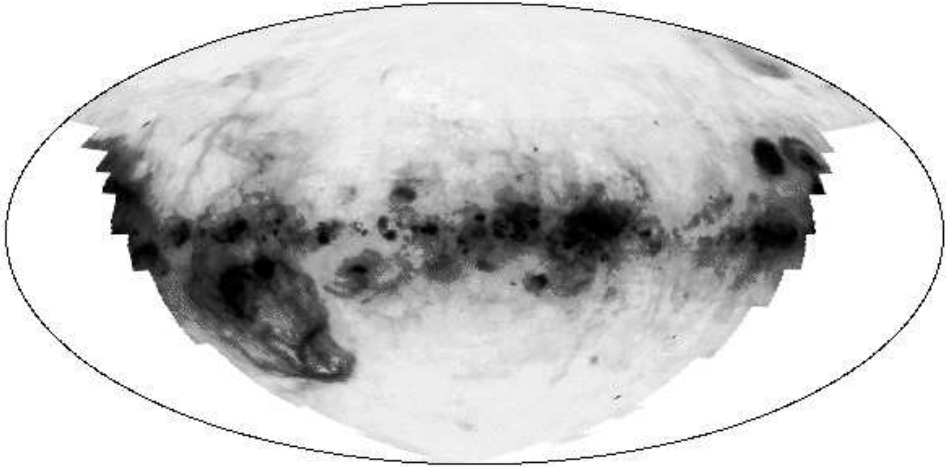,width=5.0truein}
\caption{WHAM \Ha\ survey of the Milky Way warm ionized medium,
centered at $l=120^\circ$ (Reynolds 1998; 
{\tt http://www.astro.wisc.edu/wham/}).
\label{F_wham}}
\end{figure*}

Another global effect of radiative feedback is the diffuse,
warm ionized medium (WIM).  This $10^4$ K component of the ISM
contributes $\sim 40$\% of the total \Ha\ luminosity in star-forming
galaxies for a wide variety of Hubble types (e.g., Walterbos 1998).
In the Galaxy, the WIM has a scale height of $\sim 1$ kpc, temperature
of $\sim 8000$ K, and mean density $\sim 0.025\ \cc$ (Minter \& Balser
1997).  While it has long been thought that massive stars dominate its
ionization (e.g., Frail et al. 1991; Reynolds \& Tufte 1995),
contributions from other processes also appear to be necessary.
Dissipation of turbulence (Minter \& Spangler 1997; Minter \& Balser
1997) and photoelectric heating (Reynolds \& Cox 1992)
are among the suggested heating candidates in our Galaxy. 

The WIM is most often studied through optical nebular emission.  For
the Galaxy, the largest optical survey is from the Wisconsin \Ha\
Mapper (WHAM) project (Reynolds et al. 1998; Figure~\ref{F_wham}).  In
addition to \Ha, the WHAM Fabry-Perot data also include observations of 
[S II]$\lambda$6717, [N II]$\lambda\lambda$6583, 5755, 
[O III]$\lambda$5007, He I $\lambda$5876, and other nebular emission
lines.  The other disk galaxies in the Local Group have also been
studied optically:  the LMC (Kennicutt et al. 1995), M31 (Galarza et
al. 2000; Greenawalt et al. 1997; Walterbos \& Braun 1992, 1994), M33
(Hoopes \& Walterbos 2000), and NGC~55 (Otte \& Dettmar 1999; Ferguson
et al. 1996). 

Other techniques, notably at radio wavelengths, are available for
studying the WIM in the Milky Way.  These offer additional probes of
the WIM distribution and filling factor.  Heiles et al. (1998) observed
radio recombination lines in the Galaxy, and Frail et al. (1991)
examined lines of sight through the WIM via pulsar dispersion
measures.  Faraday rotation obtained through radio polarimetry has
been exploited by e.g., Uyaniker et al. (2003), Gray et al. (1999), and
Minter \& Spangler (1996); this technique is also used by Berkhuijsen
et al. (2003) for M31. 

\section{Mechanical Feedback}

\begin{figure*}
\hspace*{0.5 truein}
\psfig{figure=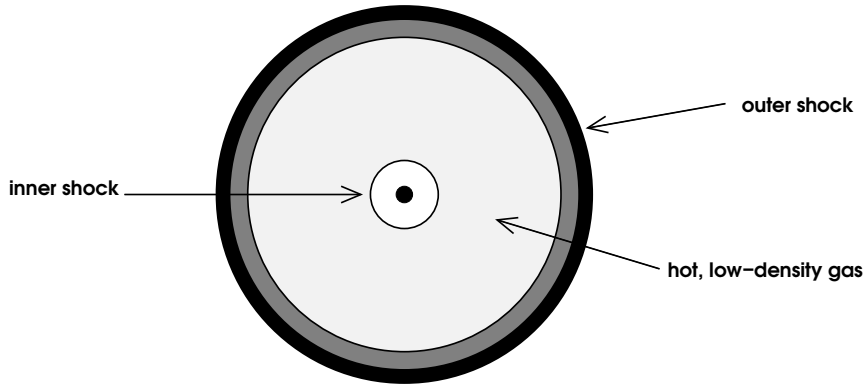,height=2.0truein}
\caption{The adiabatic wind-driven bubble model (see text).  The shell
at the outer shock consists of swept-up ISM on the outside and shocked
wind material on the inside, separated by a contact discontinuity.
\label{F_bubble}}
\end{figure*}

Supernova (SN) explosions of massive stars are a dominant source of kinetic
energy in the interstellar medium.  Strong, supersonic stellar winds
are also an important source in the case of the extreme most massive
stars ($\gtrsim 40 \rm M_\odot$).  Mechanical feedback structures the ISM,
with immediate consequences corresponding to supernova remnants
(SNRs), stellar wind-driven bubbles, and superbubbles resulting from
combined SNe and winds from multiple stars.  

While the SNRs evolve, in the simplest description, according to the
Sedov (1959) model, the wind-driven bubbles and superbubbles are
thought to evolve according to a similar,
Sedov-like adiabatic model for constant energy input (Pikel'ner
1968; Castor et al. 1977):  the central supersonic wind drives a shock
into the ambient ISM, piling up a radiatively cooled, dense shell;
and a reverse shock near the source thermalizes the wind's kinetic
energy, thereby generating a hot ($10^6 - 10^7$ K), low-density
($n\sim 10^{-2} - 10^{-3}\ \cc$) medium that dominates the bubble
volume (Figure~\ref{F_bubble}).  This heating process is believed to
be the origin of the 
diffuse hot, ionized medium (HIM) in the interstellar medium.
Assuming that the hot bubble interior remains adiabatic, the
self-similar shell evolution follows the simple analytic relations, 
\begin{eqnarray}\label{eqAD}
R & \propto & (L/n)^{1/5}\ t^{3/5} \quad , \nonumber \\
v & \propto & (L/n)^{1/5}\ t^{-2/5} \quad ,
\end{eqnarray}
where $R$ and $v$ are the shell radius and expansion velocity, $L$ is
the input mechanical power, and $t$ is the age.  Once SNe begin to
explode, they quickly dominate $L$, and the standard treatment is to
consider the discrete SNe as a constant energy input (e.g., Mac Low \&
McCray 1988).  Hence, we may write $L$ in terms of the SN parameters:
\begin{equation}\label{eq_LSN}
L=N_* E_{51}/t_e \quad ,
\end{equation}
where $N_*$ is the number of SNe, $E_{51}$ is the SN energy, and $t_e$
is the total time during which the SNe occur.

There are several approaches to testing the standard, adiabatic shell
evolution, and by extension, our understanding of mechanical
feedback.  In the first instance, we can examine the properties and
kinematics of individual shell systems and carry out rigorous
comparisons with the model predictions.  Secondly, we can also examine
statistical properties of entire shell populations in galaxies, and
compare with model predictions.  And thirdly, we can carry out spatial
correlations of shells with regions of recent star formation, to
confirm the existence of putative stellar progenitors.  All three
of these methods require high spatial resolution, and thus
the Local Group offers by far the best, and often the only feasible,
laboratory. 

\subsection{Individual shell systems}

\begin{figure*}
\hspace*{0.25 truein}
\psfig{figure=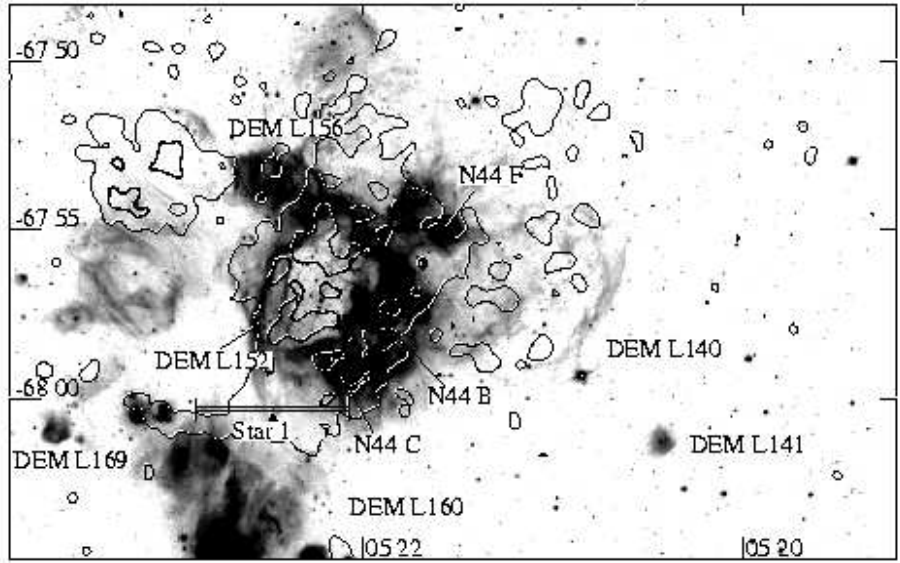,height=3.0truein}
\caption{\Ha\ image of superbubble DEM L152 (N44) in the LMC, with
{\it ROSAT} X-ray contours overlaid (from Magnier et al. 1996).
\label{F_n44}}
\end{figure*}

A number of individual superbubbles exhibit multi-phase ISM, as is
qualitatively predicted by the adiabatic shell model.  DEM
L152 (N44) in the Large Magellanic Cloud (LMC) is a beautiful example
where the nebular ($10^4$ K) gas in the shell clearly confines the
hot, X-ray--emitting ($10^6$ K) gas within (Magnier et al. 1996;
Figure~\ref{F_n44}).  Chu 
et al. (1994) also confirmed the existence of C IV and Si IV absorption
in the lines of sight toward all stars within LMC superbubbles.  These
tracers of intermediate temperature ($10^5$ K) gas are expected in
interface regions between hot and cold gas.  Quantitatively, however,
the detected X-ray emission from LMC superbubbles has been an order of
magnitude higher than predicted by the adiabatic model.  It is
therefore thought that the anomalous emission results from impacts to
the shell wall by internal SNRs (Chu \& Mac Low 1990; Wang \& Helfand
1991), a scenario which is supported by other signatures such as
enhanced [S II]/\Ha\ and anomalous kinematics (Oey 1996; see below).
Many other superbubbles have not been detected in X-rays, although the
upper limits tend to be high, and remain within the model predictions.
It is hoped that the capabilities of {\sl Chandra} and {\sl XMM} will
be applied to these objects.

Since most early-type stars are found in OB associations, wind-driven
bubbles of individual massive stars are rare, and consequently few
quantitative studies of these objects exist.  In principle, O stars
offer the most straightforward test of the standard shell evolution,
since their wind histories are simple and relatively well-understood.
One of the few such studies was carried out by Oey \& Massey (1994) on
two nebular bubbles around individual late-type O stars in M33.  They
found crude consistency with the model predictions, but the constraints
are limited by lack of kinematic information.  Cappa and collaborators
(e.g., Cappa \& Benaglia 1999; Benaglia \& Cappa 1999; Cappa \&
Herbstmeier 2000) have studied a number of \hi\ shells around Of
stars, which are presumably evolved O stars.  They generally find a
significant growth-rate discrepancy such that the shells appear to be
too small for the assumed stellar wind power.  Wolf-Rayet (W-R) stars
are well-known to have the most powerful stellar winds, and a number
of W-R ring nebulae have also been examined kinematically.  Optical
studies include those by Treffers \& Chu (1982), Garc\'\i a-Segura \&
Mac Low (1995), and Drissen et al. (1995); while, e.g., Arnal et al. (1999)
and Cappa et al. (2002) examine the neutral and radio continuum properties.
The W-R nebulae are also apparently too small, but owing to the
complicated stellar wind history and associated environment, it is
more difficult to interpret the W-R shell dynamics.

Superbubbles around OB associations are much larger, brighter, and
easier to identify than single star wind-driven bubbles, and
consequently have been much more actively studied.  They are
especially prominent in the LMC, where the proximity and high galactic
latitude offer a clear, detailed view of the objects.  The most
comprehensive study of superbubble dynamics was carried out by Oey
and collaborators (Oey 1996; Oey \& Smedley 1998; Oey \& Massey 1995)
on a total of eight LMC objects.  Saken et al. (1992) and Brown et
al. (1995) also examined two Galactic objects.  All of these are
young, nebular superbubbles having ages  $\lesssim 5$ Myr.  These
studies again consistently reveal a growth-rate discrepancy equivalent
to an overestimate in the inferred $L/n$ by up to an order of magnitude.
About half of the objects also show anomalously high expansion
velocities, implying a strong, rapid shell acceleration from the
standard evolution. 

A number of factors could be individually, or collectively,
responsible for these dynamical discrepancies.  The first possibility
is a systematic overestimate in $L/n$:   stellar wind parameters
remain uncertain within factors of 2 -- 3.  The ambient density
distribution is also critical to the shell evolution; as shown by Oey
\& Smedley (1998), a sudden drop in density can cause a
``mini-blowout'' of the shell, whose kinematics can reproduce those of
the high-velocity LMC shells.  In the case of those objects, however,
the presence of anomalously high X-ray emission favors the SNR impact
hypothesis discussed above.  In any case, the critical role of the
ambient environment motivated us to map the \hi\ distribution around
three superbubbles in the LMC sample (Oey et al. 2002).  The results
were surprisingly inhomogeneous, with one object essentially in a
void, another with significant \hi\ in close proximity, and a third
with no correspondence at all between the nebular and neutral gas.
Thus it is virtually impossible to infer the ambient \hi\ properties
for any given object without direct observations.  This heterogeneity
suggests that a systematic underestimate of $n$ is not responsible for
the universal growth-rate discrepancy.  However, a related
environmental parameter is the ambient pressure.  If the ISM pressure
has been systematically underestimated, then the superbubble growth
would become pressure-confined at an earlier, smaller stage.  We are
currently exploring this possibility (Oey \& Garc\'\i a-Segura 2003,
in preparation). 

Finally, if the superbubble interiors are somehow cooling, then
the shells will no longer grow adiabatically.  While this possibility
has been explored theoretically from several angles, there is as yet
no empirical evidence, in particular, radiation, that the objects are
cooling.  Meanwhile, mass-loading has long been a candidate cooling
mechanism (e.g., Cowie et al. 1981; Hartquist et al. 1986), either by
evaporating material from the shell wall, or by ablating clumps
that are overrun by the shell.  The enhanced density would then
increase the cooling rate of the hot interior.  More recently, Silich
et al. (2001) and Silich \& Oey (2002) suggested that the metallicity
increase expected from the parent SN explosions can significantly
enhance the cooling and X-ray emission, especially for extremely low
metallicity systems.

\subsection{Global Mechanical Feedback}

\begin{figure*}
\psfig{figure=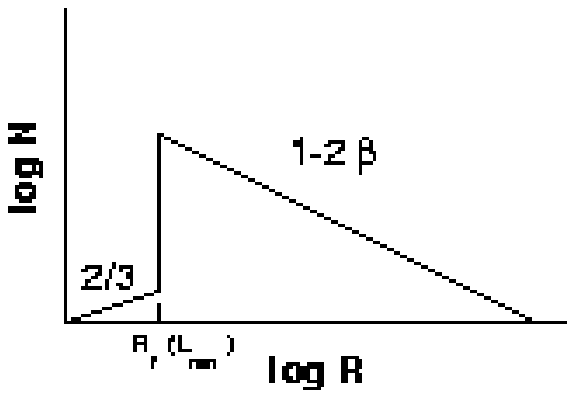,height=2.0truein}
\vspace*{-2.0truein}\hspace*{2.7truein}
\psfig{figure=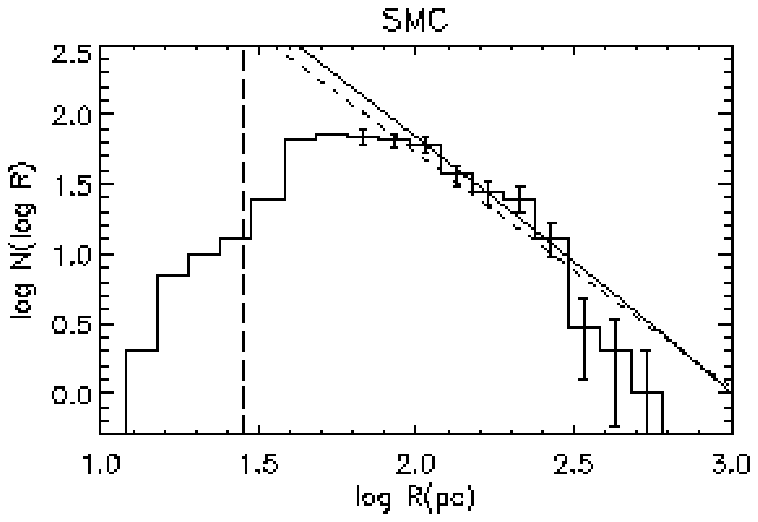,height=2.0truein}
\caption{{\it Left:}  Schematic representation of the predicted
superbubble size distribution for a power-law MLF of slope $-\beta$
and constant object creation rate.  {\it Right:}  Comparison of
slope (dotted line) fitted to the observed size distribution of \hi\
shells in the SMC and predicted slope (solid line; from Oey \& Clarke
1997).   
\label{F_size}}
\end{figure*}

Another approach to understanding superbubble evolution is in
examining the statistical properties of superbubble populations in
galaxies.  It is possible to derive the distributions in, for example,
size and expansion velocity from equations~\ref{eqAD} and assumptions
for the mechanical luminosity function (MLF), object creation rate,
and ambient parameters.  Oey \& Clarke (1997) analytically derived the
superbubble size distribution for simple combinations of creation rate
and MLF.  They found that the size distribution is dominated by
pressure-confined objects that are no longer growing, following a
differential distribution in radius $R$:
\begin{equation}\label{eq_sizes}
N(R)\ dR \propto R^{1-2\beta}\ dR \quad ,
\end{equation}
where $\beta$ is the power-law slope of the MLF $\phi(L)$ for the
form, $\phi(L) \propto L^{-\beta}$.  Since the mechanical power $L$ 
is determined by the SN progenitors (equation~\ref{eq_LSN}), the MLF
to first order has the universal power-law slope of --2 given by
equation~\ref{eq_N*}.  Equation~\ref{eq_sizes} shows that the form of
the MLF turns out to be a vital parameter in determining the size
distribution.  For the MLF slope $\beta=2$, the size distribution
$N(R)\ dR\propto R^{-3}\ dR$.  There is a peak in 
$N(R)$, corresponding to the stall radius of the lowest-$L$ objects,
which would be individual SNRs in this analysis (Figure~\ref{F_size}).

This prediction can be compared to the observed size distribution of
\hi\ shells that have been catalogued in the largest Local Group
galaxies.  By far the most complete catalog is that for the Small
Magellanic Cloud (SMC; Staveley-Smith et al. 1997).  The companion
survey for the LMC (Kim et al. 1999) surprisingly shows about four
times {\it fewer} shells than the SMC, in spite of the former's much
larger size.  The number of \hi\ shells in the LMC is also much
smaller than expected from the number of \hii\ regions, whereas their
relative numbers are fully consistent with their respective life
expectancies in the SMC (Oey \& Clarke 1997).  Thus it appears that
some process may be destroying the LMC shells prematurely, perhaps
merging, shearing in the disk, or other ISM dynamical processes
related to the high LMC star-formation rate and/or disk morphology.
In contrast, the SMC has a more 3-dimensional, solid-body kinematic
structure, and thus offers a better ISM for comparison with the crude
size distribution predictions.  

For the SMC, the observed \hi\ shell size distribution has a slope of
$-2.7\pm 0.6$, in remarkable agreement with the predicted slope of
$-2.8\pm 0.4$ derived from the actual \hii\ LF (Oey \& Clarke 1997;
Figure~\ref{F_size}). 
This suggests that the bubbly structure in the neutral ISM
of this galaxy can be entirely attributed to mechanical feedback.  It
is also worth noting that a size distribution of fractal holes can be
derived from the same dataset (Stanimirovi\'c et al. 1999).  The
power-law slope of --3.5 for the holes is similar, but different, from
that for the shells, so it will be extremely interesting to make
further such comparisons in other galaxies.

\hi\ shell catalogs also have been compiled for M31 (Brinks \& Bajaja
1986) and M33 (Deul \& den Hartog 1990).  These older catalogs lack
sensitivity and resolution, but preliminary comparisons of the shell
size distributions are broadly in agreement with the prediction (Oey
\& Clarke 1997).  A modern \hi\ survey of M33 by Thilker et al. (2000)
is also eagerly anticipated.  For the Milky Way, no complete samples
of \hi\ shells exist, but the International Galactic Plane Surveys
(IGPS; e.g., McClure-Griffiths et al. 2002) may eventually yield data
useful for a statistically significant sample.

Similarly, it is possible to derive the distribution in expansion
velocities for the shells.  Oey \& Clarke (1998b) find,
\begin{equation}
N(v)\ dv \propto v^{-7/2}\ dv \quad , \quad \beta > 3/2 \quad .
\end{equation}
Again, comparison with the SMC catalog shows consistency with the prediction,
although with much larger uncertainty in the observed slope of $-2.9\pm
1.4$.  Although taken from the same dataset, this comparison examines
a different subset of objects, since the shells with non-negligible
expansion velocities clearly sample the growing objects, whereas the
size distribution is dominated by pressure-confined, stalled objects.

\begin{figure*}
\hspace*{0.2truein}
\psfig{figure=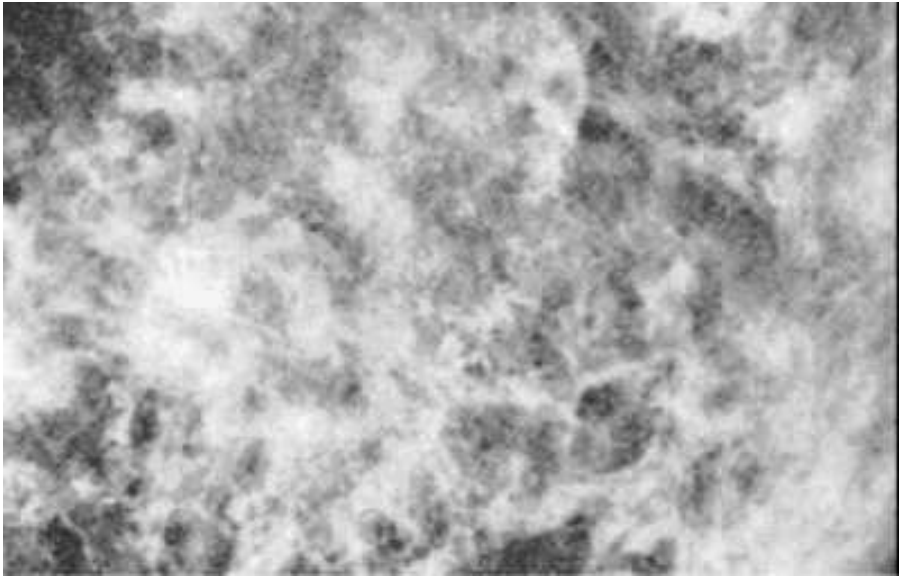,height=3.0truein}
\caption{\hi\ distribution in a central region of the LMC (white is
positive), showing a filamentary ISM morphology that is inconsistent
with simple fractal models (from Elmegreen et al. 2001).
\label{F_LMCmorph}}
\end{figure*}

The Local Group galaxies, especially the Magellanic Clouds, also
provide an opportunity to examine the detailed morphology of the ISM.
An especially compelling discussion is presented by Elmegreen et
al. (2001), who compare the structural morphology of the neutral ISM
in the LMC with a fractal model.  They show that the LMC's ISM is
clearly more filamentary than accounted for in their simple fractal
model (Figure~\ref{F_LMCmorph}).  Such filamentary structure can be
attributed at least in part 
to the superbubble structuring caused by mechanical feedback.  Oey
(2002) reviews \hi\ structure in the ISM and
argues that the origin of filamentary structure is key to
understanding the dynamical and evolutionary processes in the ISM,
such as phase balance and star formation.

\subsection{Correspondence with Star Formation}

Modern studies that compare the spatial distribution of \hi\ shells and
star-forming regions show perhaps surprisingly ambiguous results.  The
galaxy that has been most actively studied in this respect recently is
Holmberg~II.  At a distance of 3~Mpc, Ho~II is not a member of the
Local Group, and the conflicting results in the literature may be
symptomatic of the difficulty with spatial resolution at that distance.
While Rhode et al. (1999) failed to identify \hi\ shell progenitor
populations from $BVR$ aperture photometry, Stewart et al. (2000) did
find a positive spatial correlation with star-forming regions using
FUV data from the {\sl Ultraviolet Imaging Telescope} ({\sl UIT})and
\Ha\ observations.  Tongue \& Westpfahl (1995) also found that the SN
rate implied by the radio continuum emission is consistent with the
total, integrated superbubble energy in Ho~II.

The Magellanic Clouds, and other nearby Local Group galaxies, are
clearly superior candidates for investigating quantitative spatial
correlations.  Kim et al. (1999) have carried out a preliminary study
that compares the \hi\ shell properties with those of the OB
associations and nebular emission.  They are able to identify an
evolutionary sequence such that the shells with associated \Ha\
emission show higher expansion velocities than those showing only the
presence of OB stars, and the latter in turn show higher $v$ than the
remainder of the shells.  The \Ha\ emission also shows smaller radial
extent, remaining within that of the \hi\ shells.  These trends are
consistent with an age sequence and feedback origin for the neutral
shells.  We are currently carrying out a more detailed follow-up of
this study using {\sl UIT} and new optical data (Oey et al. 2003, in
preparation). 

\subsection{Interstellar Porosity}

\begin{figure*}
\hspace*{0.6 truein}
\psfig{figure=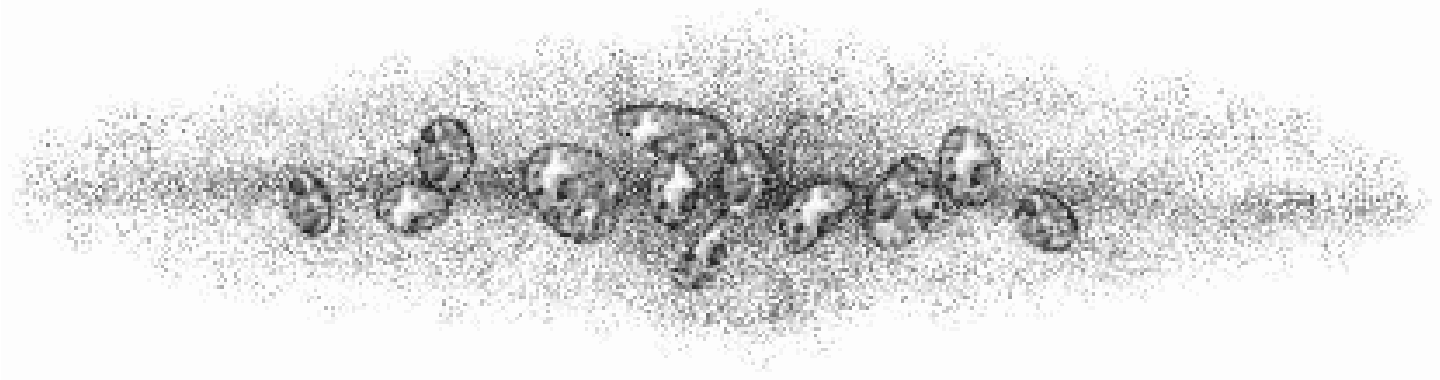,width=4.0truein}

\vspace*{-0.8 truein}
\hspace*{0.6 truein}
\psfig{figure=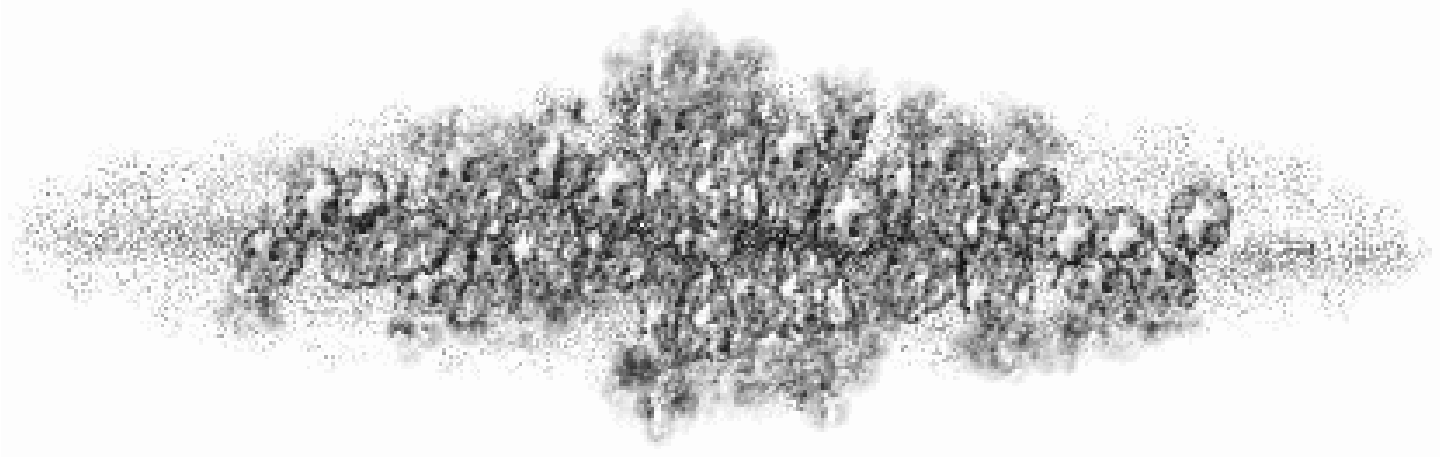,width=4.0truein}
\caption{Low and high interstellar porosities are shown schematically
in the upper and lower panels, respectively.  The lower panel shows
how $Q = 1$ defines a threshold porosity and star-formation rate for
galactic outflows and escape of ionizing radiation.
\label{F_fdbk}}
\end{figure*}

A conventional parameterization for global mechanical feedback in
galaxies is the interstellar porosity, or volume filling factor of 
superbubbles.  Having derived analytic expressions for the superbubble
size distribution, it is straightforward to derive expressions for the
porosity (Oey \& Clarke 1997).  This can be written in terms of the
star formation rate (SFR), galactic scale height $h$, and galactic
star-forming radius $R_g$ (Oey et al. 2001):
\begin{equation}\label{eqQSFR}
Q \simeq 16 \frac{{\rm SFR}({\rm M_\odot\ yr^{-1}})}{hR_g^2(\rm kpc^3)}
	\quad ,
\end{equation}
for Milky Way ISM parameters and Salpeter (1955) IMF.  The porosity
also can be written in terms of galactic parameters, i.e., ISM mass
and velocity dispersion (Clarke \& Oey 2002).  Assuming a feedback
origin for the HIM, the interstellar porosity quantifies the relative
phase balance between the HIM and cooler ISM phases.  $Q$ greater
than a critical value of unity therefore indicates an outflow
condition for the hot gas, as might be encountered in a starburst
situation (Figure~\ref{F_fdbk}). 

For the Local Group star-forming galaxies, Oey et al. (2001) find
$Q\ll 1$ in almost all cases, suggesting that the HIM generally does
not dominate the ISM volume.  The LMC, however, does show $Q\sim 1$.
The Milky Way situation is ambiguous, with different estimates of the
SFR yielding $Q$ in the range 0.2 to 1.  The truly glaring exception is
the starburst galaxy IC~10, for which $Q \gtrsim 20$, clearly
fulfilling the outflow or superwind criterion.  Clarke \& Oey (2002)
also posit the critical $Q = 1$ condition as a threshold for radiative
feedback to the IGM.  Once mechanical outflow is established, the
merging and blowout of superbubbles also opens free pathways for
ionizing photons to escape from the parent galaxy.  They apply this
simple model to a variety of phenomena, ranging from giant molecular
clouds to Lyman break galaxies.

\section{Conclusion}

It is clear that both radiative and mechanical feedback energies 
can dominate evolutionary processes in star-forming galaxies,
although we still lack understanding of important aspects in the
feedback processes.  The extreme range in scale over which feedback
has influence is especially remarkable.  In addition, it is also
possible to investigate and parameterize chemical feedback with
analogous methods (e.g., Oey 2000, 2003).  

Radiative feedback is responsible for the nebular
emission-line diagnostics and tracers of star formation, as well as
the WIM component of the ISM.  On a detailed level, the line
diagnostics still need improved calibrations and photoionization
modeling to increase their utility and parameter space.  The \hii\ LF
is emerging as a quantitative diagnostic of global star formation in
galaxies, and reveals a universal law in the stellar membership
function for massive stars (equation~\ref{eq_N*}).  The ionization and
energy budget of the WIM, while apparently tied to massive stars,
remains to be better understood.  Ultimately, the escape of ionizing
radiation beyond the parent galaxies bears upon the intergalactic
environment and reionization of the early Universe.

Mechanical feedback indisputably drives shell structures that are
ubiquitous in the ISM:  SNRs, stellar wind-driven bubbles, and
superbubbles.  The standard, adiabatic model for the evolution of the
bubbles and superbubbles is broadly consistent with a variety of
empirical evidence; yet, quantitatively, a number of outstanding
problems remain.  The late evolution of the shells is especially
enigmatic and critical for understanding the role of feedback in
generating the HIM, as well as the global properties of the ISM.
Spatial correlations of interstellar shells with star-forming regions
still need to firmly establish the physical processes and interactions
related to feedback.  Preliminary studies of the size and velocity
distributions of \hi\ shells appear to confirm a dominant role for
mechanical feedback in structuring the ISM.  This analytic analysis
can be extended to parameterize the interstellar porosity, implying a
critical threshold for the outflow of superwinds, heavy elements, and
ionizing radiation.

\begin{acknowledgments}
I am grateful to the conference organizers and STScI for supporting
in part this contribution to the Symposium.
\end{acknowledgments}

\vfill\noindent
{\it  The Local Group as an Astrophysical Laboratory,} 2003 STScI May
Symposium, eds. M. Livio et al., (Cambridge:  Cambridge U. Press)

\end{document}